\documentclass[11pt]{cernart}
\usepackage[dvips]{graphics,rotating,epsfig} \usepackage{amsmath}
\usepackage{times} \usepackage{xspace} \usepackage{array}

\newcommand{\dch}{\ensuremath{D_{\mathrm{ch}}}\xspace}
\newcommand{\nch}{\ensuremath{n_{\mathrm{ch}}}\xspace}
\newcommand{\nbl}{\ensuremath{n_{\mathrm{g}}}\xspace}
\newcommand{\nsh}{\ensuremath{n_{\mathrm{s}}}\xspace}
\newcommand{\ngr}{\ensuremath{n_{\mathrm{g}}}\xspace}
\newcommand{\nh}{\ensuremath{n_{\mathrm{h}}}\xspace}
\newcommand{\wsq}{\ensuremath{W^{2}}\xspace}
\newcommand{\npls}{\ensuremath{n_{\mathrm{+}}}\xspace}
\newcommand{\nneg}{\ensuremath{n_{\mathrm{-}}}\xspace}
\newcommand{\Enu}{\ensuremath{E_{\mathrm{\nu}}}\xspace}
\newcommand{\Enubar}{\ensuremath{E_{\mathrm{\overline{\nu}}}}\xspace}
\newcommand{\Qsq}{\ensuremath{Q^{\mathrm{2}}}\xspace}
\newcommand{\Qsqnu}{\ensuremath{\Qsq_{\mathrm{\nu}}}\xspace}
\newcommand{\Qsqnubar}{\ensuremath{\Qsq_{\mathrm{\overline{\nu}}}}\xspace}
\newcommand{\wsqnu}{\ensuremath{\wsq_{\mathrm{\nu}}}\xspace}
\newcommand{\wsqnubar}{\ensuremath{\wsq_{\mathrm{\overline{\nu}}}}\xspace}

\newcommand{\meantheta}{\ensuremath{\langle \theta \rangle}\xspace}
\newcommand{\meannpls}{\ensuremath{\langle \npls \rangle}\xspace}
\newcommand{\meannneg}{\ensuremath{\langle \nneg \rangle}\xspace}

\newcommand{\meanwsq}{\ensuremath{\langle \wsq \rangle}\xspace}
\newcommand{\meannch}{\ensuremath{\langle \nch \rangle}\xspace}
\newcommand{\meannbl}{\ensuremath{\langle \nbl \rangle}\xspace}
\newcommand{\meanEnu}{\ensuremath{\langle \Enu \rangle}\xspace}
\newcommand{\meanEnubar}{\ensuremath{\langle \Enubar \rangle}\xspace}

\newcommand{\meanQsq}{\ensuremath{\langle \Qsq \rangle}\xspace}
\newcommand{\meanQsqnu}{\ensuremath{\langle \Qsqnu \rangle}\xspace}
\newcommand{\meanQsqnubar}{\ensuremath{\langle \Qsqnubar \rangle}\xspace}

\newcommand{\numu}{\ensuremath{\nu_\mu}\xspace}
\newcommand{\numubar}{\ensuremath{\overline{\nu}_\mu}\xspace}

\newcommand{\nuA}{\ensuremath{\nu\mbox{--A}}\xspace}
\newcommand{\nubarA}{\ensuremath{\overline{\nu}\mbox{--A}}\xspace}

\newcommand{\kg}{\ensuremath{\mbox{kg}}\xspace}
\newcommand{\GeV}{\ensuremath{\mbox{GeV}}\xspace}

\newcommand{\GeVc}{\ensuremath{\mbox{GeV}/c}\xspace}
\newcommand{\GeVcsq}{\ensuremath{(\mbox{GeV}/c)^2}\xspace}

\newcommand{\GeVccsq}{\ensuremath{\mbox{GeV}^2/c^4}\xspace}

\newcommand{\cm}{\ensuremath{\mbox{cm}}\xspace}

\newcommand{\micron}{\ensuremath{\mu \mbox{m}}\xspace}

\newcommand{\mrad}{\ensuremath{\mbox{mrad}}\xspace}

\begin{document}
\begin{titlepage}

\rightline {CERN-PH-EP/2007-023}
\rightline {09 July 2007}
\vglue 0.1cm

\title{\Large \bf Charged-Particle Multiplicities in Charged-Current
  Neutrino-- and Anti-Neutrino--Nucleus Interactions }

\author{CHORUS Collaboration}
\maketitle

\begin{center}
\end{center}

\begin{abstract}

The CHORUS experiment, designed to search for $\nu_{\mu}\rightarrow\nu_{\tau}$ oscillations, consists of a nuclear emulsion
target and electronic detectors. In this paper,
results on the
production of charged particles  in a small sample of
charged-current neutrino--
and anti-neutrino--nucleus interactions at high energy are presented.
For each event, the emission angle and the ionization
features of the charged particles produced in the
interaction are recorded, while the standard kinematic
variables are reconstructed using the electronic detectors.
The average multiplicities for charged tracks, the pseudo-rapidity
distributions, the dispersion in the multiplicity of charged particles
and the KNO scaling are studied in different kinematical
regions. 
A study of quasi-elastic topologies performed for the first time in
nuclear emulsions is also reported.
The results are presented in a form suitable for use in the validation of
 Monte Carlo generators of neutrino--nucleus interactions.

\begin{center}
\end{center}

\end{abstract}

\vspace{5.0cm}
\submitted{To be published in European Physical  Journal C}

\newpage

\begin{center}   
{\Large {CHORUS Collaboration}}
\end{center}
\vspace{0.1cm}

\begin{Authlist}
A.~Kayis-Topaksu, G.~\"{O}neng\"ut

{\bf \c{C}ukurova University, Adana, Turkey}

R.~van Dantzig,  M.~de Jong, 
R.G.C.~Oldeman$^1$

{\bf NIKHEF, Amsterdam, The Netherlands}

M.~G\"uler, U.~K\"ose, 
P.~Tolun

{\bf  METU, Ankara, Turkey}

M.G.~Catanesi, 
M.T.~Muciaccia

{\bf Universit\`a di Bari and INFN, Bari, Italy}

K.~Winter

{\bf  Humboldt Universit\"at, Berlin, Germany$^{2}$}

B.~Van de Vyver$^{3,4}$, P.~Vilain$^{5}$, G.~Wilquet$^{5}$

{\bf Inter-University Institute for High Energies (ULB-VUB) Brussels, Belgium}

B.~Saitta

{\bf Universit\`a di Cagliari and INFN, Cagliari, Italy}

E.~Di Capua

{\bf Universit\`a di Ferrara and INFN, Ferrara, Italy}

S.~Ogawa, H.~Shibuya

{\bf Toho University,  Funabashi, Japan}

I.R.~Hristova$^6$, 
T.~Kawamura,
D.~ Kolev$^7$, H.~ Meinhard, 
J.~Panman, 
A.~Rozanov$^{8}$,
R.~Tsenov$^{7}$, J.W.E. Uiterwijk, P. Zucchelli$^{3,9}$

{\bf CERN, Geneva, Switzerland}

J.~Goldberg

  {\bf Technion, Haifa, Israel}

M.~Chikawa

 {\bf Kinki University, Higashiosaka, Japan}

J.S.~Song, C.S.~Yoon

{\bf Gyeongsang National University,  Jinju, Korea}

K.~Kodama, N.~Ushida

{\bf Aichi University of Education, Kariya, Japan}

S.~Aoki, T.~Hara

 {\bf Kobe University,  Kobe, Japan}

T.~Delbar,  D.~Favart, G.~Gr\'egoire, S.~ Kalinin, I.~ Makhlioueva

{\bf Universit\'e Catholique de Louvain, Louvain-la-Neuve, Belgium} 

A.~Artamonov, P.~Gorbunov, V.~Khovansky, V.~Shamanov, I.~Tsukerman

{\bf Institute for Theoretical and Experimental Physics, Moscow, Russian
Federation}

N.~Bruski, D.~Frekers

{\bf Westf\"alische Wilhelms-Universit\"at, M\"unster, Germany$^{2}$}

K.~Hoshino, 
J.~Kawada, 
M.~Komatsu,
M.~Miyanishi, 
M.~Nakamura, T.~Nakano, K.~Narita, K.~Niu, K.~Niwa, 
N.~Nonaka, O.~Sato, T.~Toshito

{\bf Nagoya University, Nagoya, Japan}

S.~Buontempo, A.G.~Cocco, N.~D'Ambrosio,
G.~De Lellis, G.~ De Rosa, F.~Di Capua, 
G.~Fiorillo, A.~Marotta,
P.~ Migliozzi, 
L.~Scotto Lavina, 
P.~ Strolin, V.~Tioukov

{\bf Universit\`a Federico II and INFN, Naples, Italy}

T.~Okusawa

{\bf Osaka City University, Osaka, Japan}

U.~Dore, P.F.~Loverre,
L.~Ludovici, 
G.~Rosa, R.~Santacesaria, A.~Satta, F.R.~Spada

{\bf Universit\`a La Sapienza and INFN, Rome, Italy}

E.~Barbuto, C.~Bozza, G.~Grella, G.~Romano, C.~Sirignano, S.~Sorrentino

{\bf  Universit\`a di Salerno and INFN, Salerno, Italy}

Y.~Sato, I.~Tezuka

{\bf Utsunomiya University,  Utsunomiya, Japan}

{\footnotesize
---------

\begin{flushleft}

$^{1}$ {Now at Universit\`a La Sapienza, Rome, Italy.}
\newline
$^{2}$ {Supported by the German Bundesministerium f\"ur Bildung und Forschung under contract numbers 05 6BU11P and 05
7MS12P.}
\newline
$^{3}$ {Now at SpinX Technologies, Geneva, Switzerland.}
\newline
$^{4}$ {Fonds voor Wetenschappelijk Onderzoek, Belgium.}
\newline
$^{5}$ {Fonds National de la Recherche Scientifique, Belgium.}
\newline
$^{6}$ {Now at DESY, Hamburg.}
\newline
$^{7}$ {On leave of absence and at St. Kliment Ohridski University of Sofia, Bulgaria.}
\newline
$^{8}$ {Now at CPPM CNRS-IN2P3, Marseille, France.}
\newline
$^{9}$ {On leave of absence from INFN, Ferrara, Italy.}
\end{flushleft}
}

\end{Authlist}

\end{titlepage}

\newpage


\section{Introduction}

In the study of multi-particle production processes, the
multiplicity of charged particles is an important
global parameter reflecting the dynamics of the interaction. 
Different phenomenological and theoretical models can be
 tested and therefore multiplicities have been studied in experiments with  different particle
beams, making use of  various techniques and over a wide range of kinematic variables.

The characteristics of charged-particle multiplicity
distributions have been studied in detail in high energy hadronic 
collisions \cite{biebl}, in $e^{-}e^{+}$ annihilation
\cite{Berger,Brandelik}, and in interactions of neutrinos and anti-neutrinos on
nucleons and light
nuclei~\cite{Allen,Bell,Jones,Grassler,Zieminska}. In particular, Bubble Chamber 
experiments~\cite{Baranov,Derrick,Barlag}
performed measurement of the charged particle multiplicity, produced in \numu~(\numubar)n and \numu~(\numubar)p interactions.
They also studied the multiplicity moments as function of kinematical quantities.

The sub-micron spatial resolution of 
nuclear emulsion 
allows both
the investigation of the event topology and the measurement of the angular
distribution of charged particles to be performed with high accuracy. Therefore, 
it is well suited to study particle-production multiplicities.
However, such data on neutrino interactions in nuclear emulsion are
relatively  scarce~\cite{Voyvodik, Ammosov} and don't exist for
anti-neutrino interactions in nuclear emulsion.

In this paper, we report on a study of charged-particle
multiplicities produced in high-energy charged-current
(anti-)neutrino interactions in a nuclear emulsion target measured
with the CHORUS hybrid detector. 
Such data are useful in tuning interaction models in Monte Carlo 
 event generators. We also
investigate the KNO scaling~\cite{Koba}
behaviour in different kinematical regions.

\section{The experimental apparatus}

%
The CHORUS detector~\cite{chorus} is a hybrid setup that combines a nuclear emulsion
target with several electronic detectors: trigger hodoscopes, a  
scintillating fibre tracker system, a hadron spectrometer, a
calorimeter and a muon spectrometer.

The target  consists of {\it Fuji ET-7B}~\cite{chorusem} nuclear 
emulsions with properties given in
Table~\ref{tab:compo}. The nominal sensitivity is 30
grains per 100 \micron for minimum-ionizing particles. The target
is segmented into four stacks and has an overall mass of 770~\kg. Each
of the stacks consists of eight modules of 36 plates of size 36~\cm
$\times$ 72 \cm.
Each plate has a 90 \micron plastic support coated
on both sides with a 350 \micron emulsion layer.
Particle track directions are nearly perpendicular to the emulsion sheets.
This configuration permits fast automatic scanning of the emulsion sheets.

Each target stack is
followed by three interface emulsion sheets with  90 \micron   
emulsion layers on both sides of an 800 \micron thick plastic base
and by a set of scintillating-fibre tracker
planes~\cite{fibretrackers}.
The
interface sheets and the fibre trackers provide accurate particle
trajectory measurements which are extrapolated into the emulsion stack in order to locate
the neutrino interaction vertices.  The accuracy of the fibre tracker
prediction is about 150 \micron in the track position and 2 \mrad in angle.

The electronic detectors downstream of the emulsion target include
a hadron spectrometer that  measures the bending of charged
particles in an air-core hexagonal magnet~\cite{hexmagnet}, a
calorimeter where the energy
and direction of showers are measured and a muon spectrometer   
which measures the charge and the momentum of muons.
The energy resolution for hadronic showers of the calorimeter  is
$  \sigma(E)/E = (0.323 \pm 0.024)/\sqrt{E/\GeV}+(0.014 \pm 0.007)\
  $~\cite{calo}.
The muon spectrometer consists of six magnetized iron toroids
sandwiched between seven tracking stations (drift chambers and
streamer tubes).
The muon momentum resolution varies from
$\approx$15\% in the 12--28~\GeVc range~\cite{chorusspec}
to 19\% at about 70~\GeVc~\cite{chorus}, as measured with
test-beam muons.

The West Area Neutrino Facility (WANF) of the CERN SPS provided an
intense beam of neutrinos with an average energy of 27~\GeV. The  
beam consisted mainly of \numu with a contamination of
$\sim$5\% \numubar and $\sim$1\% $\nu_e$. The CHORUS
detector was exposed to the wide band neutrino beam (section 5 of
~\cite{chorus} for details) of the CERN 
SPS during the years 1994--1997, with an integrated flux of      
$5.06\times$ $10^{19}$ protons on target. In this four year
exposure more than $10^{6}$ neutrino interactions were accumulated
in the emulsion target.

\begin{table*}[tb]
\begin{center}
\caption{Atomic composition and main features of the nuclear emulsions (Fuji ET-B7) used
in the CHORUS experiment~\cite{Niwa}} 
\label{tab:compo}
\label{table:composition}
\vspace{0.5\baselineskip}
\newcommand{\mph}{\hphantom{$-$}}
\newcommand{\cc}[1]{\multicolumn{1}{c}{#1}}
\renewcommand{\tabcolsep}{2pc} 
\begin{tabular}{@{}llll}
\hline
 {\bf Element} & \cc{\bf Atomic number} & \cc{\bf Mass ($\%$)} &
 \cc{\bf Mole fraction ($\%$)} \\
\hline
  Iodine (I)   & \mph53            & \mph0.3           & \mph0.06                 \\
  Silver (Ag)  & \mph47            & \mph45.5          & \mph11.2                 \\
  Bromine (Br) & \mph35            & \mph33.4          & \mph11.1                 \\
  Sulphur (S)   & \mph16            & \mph0.2           & \mph0.2                  \\
  Oxygen (O)   & \mph8             & \mph6.8           & \mph11.3                 \\
  Nitrogen (N) & \mph7             & \mph3.1           & \mph5.9                  \\
  Carbon (C)   & \mph6             & \mph9.3           & \mph20.6                 \\
  Hydrogen (H) & \mph1             & \mph1.5           & \mph40.0                 \\
\end{tabular}\\[2pt]
\begin{tabular}{@{}lll}
\hline
 Mean number of nucleons &36 protons, 45 neutrons &\\
 Density &3.73$\mbox{g/cm}^3$  & \\
 Radiation length&2.94 \cm &\\
 Nuclear interaction mean free path&38 \cm&\\
 Concentration of AgBr&45.5\% in volume &\\
\hline
\end{tabular}\\[2pt]
\end{center}
\end{table*}
 
\section{Analysis}

The emulsion scanning has been performed by fully automatic
microscopes each equipped with a CCD camera and a fast custom built
processor (originally called 'Track Selector')~\cite{Nakano}  capable of
identifying tracks inside the emulsions, and measuring on-line their
parameters. The track finding efficiency of the Track
Selector is higher than 98\% for tracks with an angle of less than
400~\mrad~\cite{murat} with respect to the beam direction, perpendicular to the emulsion plates.

The location of the emulsion plate containing
the interaction vertex is achieved with a procedure which is  called scan-back~\cite{umut}. As a first step, tracks
reconstructed by the electronic detectors are followed upstream into
the emulsion stacks. 
A track, extrapolated from the fibre-trackers and found in the interface
emulsion sheets, is followed upstream in the target emulsion stack 
using track segments reconstructed in the most upstream 100~\micron
of each plate. If the track is not found in
two subsequent plates, the first plate with no track hit  is
defined as the plate containing the vertex. 
In total, about 150~000 \numu~CC events with at least one reconstructed muon in spectrometer,
have been located in emulsion as a result of this procedure. 

For the present measurement, from the 150000 located events with at 
least one reconstructed muon in the spectrometer, a sample of 1208
events was randomly selected to be visually inspected and to
measure the parameter associated with the charged tracks at the (anti-)neutrino
vertex. We use the sign of the charge of the highest-energy muon in the events, as
determined by the muon spectrometer, to determine whether the
interaction was induced by a neutrino or an anti-neutrino. In order to have a larger anti-neutrino
sample, events located in various  modules of the emulsion stacks have
been collected, while the neutrino sample consists of 
events located in one half module of an emulsion stack. In 627~(581) of the selected events the charge of the leading muon is negative~(positive). 

To evaluate the number of 
genuine  \numubar charged-current~(CC) interactions, one has to subtract  from this sample the 
contamination due to:

\begin{itemize}
\item \numu CC events with the $\mu^{-}$ reconstructed as a $\mu^{+}$;
\item punch-through hadrons traversing the 5.2 interaction lengths of
the calorimeter and reconstructed as positive muons in the
spectrometer. This can happen in CC events where the  $\mu$ is not
identified or in NC interactions;
\end{itemize}

\noindent
In order to reduce this background, a set of selection criteria
improving the reliability of the muon reconstruction is applied to the
events with a muon reconstructed with positive charge:

\begin{itemize}
\item low momentum muons stopping inside the spectrometer, for
  which the reconstruction is based on only a few hits in the electronic
  detectors, are discarded;

\item the impact point on the spectrometer
  entrance surface is required to be within a 150~\cm radius centered
  around its axis to ensure a long potential path through the toroidal 
  magnets; 

\item 
  the track length is required to be greater than three spectrometer
  gaps (two magnets) and consistency checks are made among the various
  reconstruction algorithms;

\item the trajectory of the muon measured in the spectrometer is
  required to match with a track detected in the
  fibre trackers, and a selection criterion is applied on the basis  of $\chi^{2}$ of
  the global track fit.

\end{itemize}

These criteria select 529 events with a reconstructed positive muon.
The contribution from $\pi$ and $K$ decays into a $\mu$,
reconstructed in the spectrometer is negligible, as well as \numu
induced charm events with the $\mu^{-}$ reconstructed as a $\mu^{+}$.
The remaining background to the sample of  $\mu^{+}$  due to \numu
events is composed of punch-through tracks of pions and kaons and amounts to 
$5.5 \pm 1.9 \ \mbox{(stat)} \pm 0.6 \ \mbox{(syst)}$ events.
%

We further require that square of the invariant mass of the hadronic system,
\wsq, of  (anti-)neutrino  events is  greater than 1~\GeVccsq. 
After this selection
the  number of \numu--A (\numubar--A) interactions is 496 (369).

\section{Measurement procedure and charged prong classification in nuclear emulsions}


When a charged particle passes through nuclear emulsion it  forms
a latent image by ionizing the AgBr crystals along its trajectory.
After a suitable chemical processing  of the emulsions such ionization
sites induce the deposition of  Ag atoms as dark spots (grains) thus
making the trajectory of the charged particles visible. 
Tracks are usually classified according to their
grain density (number of Ag grains per unit length) as shower,
grey and black prongs following the procedure described in
Refs.~\cite{Powell,Barkas}.
Shower tracks correspond to relativistic (or nearly  relativistic) singly charged particles with
a velocity $\beta\geq 0.7$. Their grain density is $g<1.4 \ g_{0}$ ( $g_{0}$
represents the plateau grain density of a highly  relativistic singly charged particle, which
equals 29.6 grains per 100 \micron for the CHORUS emulsion). 
Grey tracks correspond
to charged particles with velocity $0.25\leq\beta<0.7$. They are
interpreted as recoil nucleons emitted during the nuclear cascade. 
Their grain
density is $1.4 \ g_{0}<g<10 \ g_{0}$. The black prongs correspond to
 particles with a velocity $\beta <0.25$.
These are
produced by low energy fragments (protons, deuterons, alphas and
heavier fragments) emitted from the excited target nucleus. The
associated grain density is $g \ge 10 \ g_{0}$. 

The above criteria commonly
used in nuclear emulsion experiments are very difficult to
apply in the CHORUS analysis since the emulsion sheets were
exposed perpendicular to the beam direction. Furthermore, electronic measurement of the energy and
momentum for every charged particle was not possible due to
geometrical  acceptances.

\begin{figure}[tb]
\begin{center}
\resizebox{0.85\textwidth}{!}{
\includegraphics{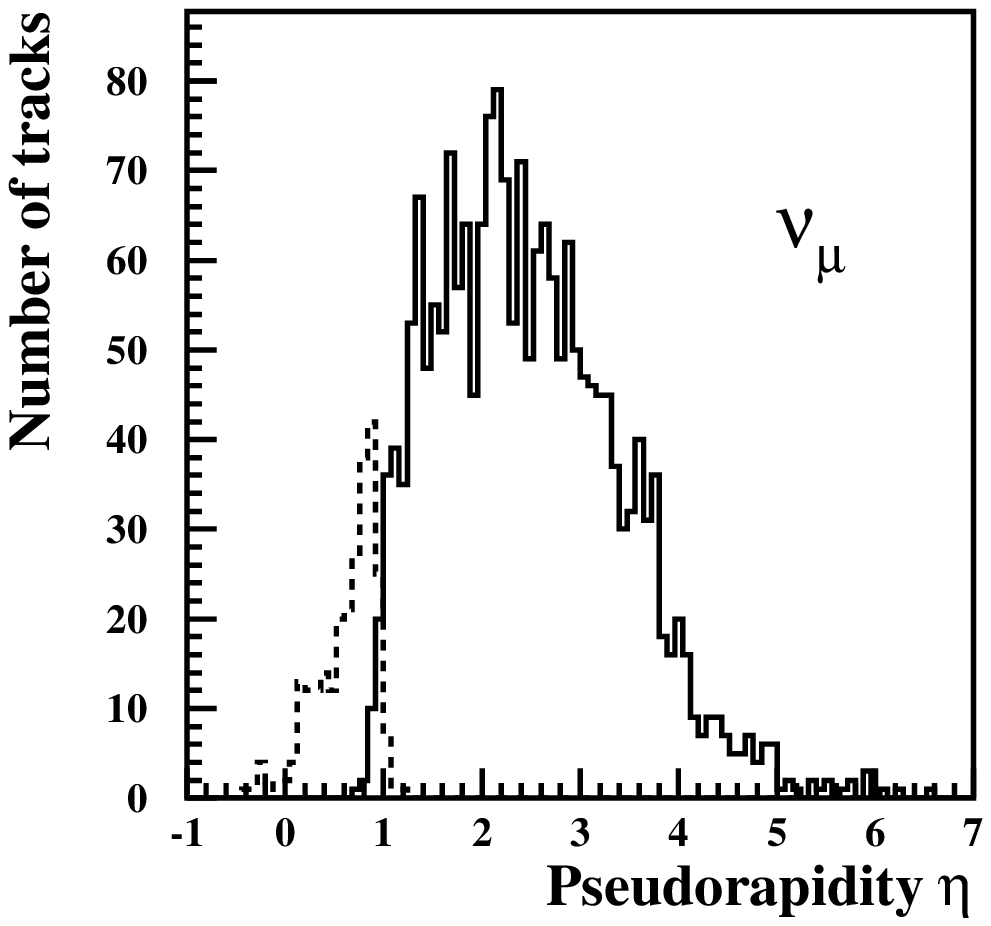}
\includegraphics{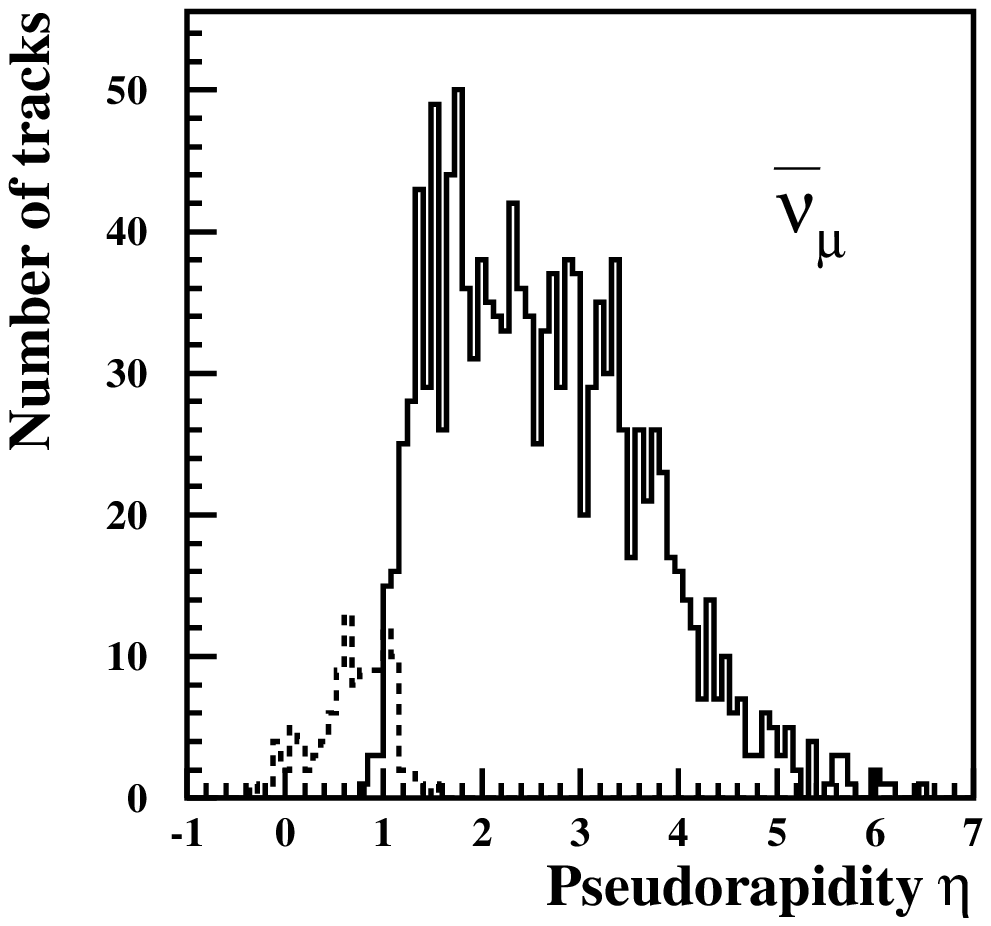} }

\caption{Pseudo-rapidity distributions for tracks classified as shower
(full histogram) and  grey 
  (dashed histogram)  by the scanner.
}
\label{fig:psedo}
\end{center}
\end{figure}
The following 
procedure was used for the CHORUS data. Once the vertex plate had been located by the 
automatic system, the event was checked and measured extensively
with a manually controlled microscope system. The black prongs
at a neutrino interaction vertex,  due to heavily ionizing particles,
have short path lengths and usually stop within one emulsion
plate so that they can be recognized.
For the remaining   mixture of shower
and grey prongs, we measured the particle directions. These particles
are mainly pions with a small contamination of protons.

Ambiguity in classifying shower and grey prongs
can arise either due to the difficulty in strictly applying the ionization criteria in the manual scanning
 or due to variations in the quality of the  optics
of the microscope. To overcome these limitations we 
decided to classify these prongs by  using the pseudo-rapidity variable~\cite{liu}: 
\begin{equation}\label{2a}
  \eta= -\ln \tan\frac{\theta}{2}
\end{equation}
where $\theta$ is the emission angle of the prong with respect to
the neutrino direction. This has the advantage of being independent of
the scanner and of the microscope optics, allowing us to compare in
a straightforward manner  the  multiplicity measurements with the theoretical models.
The pseudo-rapidity distributions for tracks classified
as shower and  grey by the 
scanner, both for neutrino and anti-neutrino interactions are shown in
 Fig.~\ref{fig:psedo}. These plots show that a scanner-independent 
classification, based on the pseudo-rapidity, is possible and it is consistent with the traditional one. 
In the following, all prongs with  $\eta \ge 1$ are classified as shower   
particles. By applying this cut the fraction of shower (grey) tracks
miss-identified as grey (shower) is 
(1.23$\pm$0.19)$\%$~((7.27$\pm$1.40$)\%$).
The multiplicities of shower, grey and heavy~(grey+black)
prongs are denoted by \nsh, \ngr  and \nh, respectively. The
total number of charged hadrons classified as shower particles in an
event is defined to be  $\nch = \nsh - 1$, namely the number of shower
tracks minus the muon track. 
\section{Efficiency Evaluation}

Reconstruction and location efficiencies were evaluated making a
detailed simulation of the detector response using  a program
based on GEANT3~\cite{geant}. Large samples of
deep-inelastic neutrino interactions (DIS) were generated using
the beam spectrum from the CHORUS JETTA generator~\cite{jetta} based on
LEPTO~\cite{lepto} and JETSET~\cite{jetset}.  Quasi-elastic (QE)
interactions and resonance production (RES) events were generated with the
RESQUE \cite{resque} package with a rate of 9.6\% relative to
deep-inelastic scattering reactions in the neutrino case (26\%
in the anti-neutrino case). The simulated response of the electronic
detectors was processed through the same reconstruction program 
as that used for the experimental data. The location efficiency was parametrized by a
 function of the primary muon momentum and angle on which it depends only weakly.
 The reconstruction and location
efficiency as a function of the invariant mass of the hadronic system \wsq and
of the shower prong multiplicities {\nch} is given in Tables~\ref{tab:nueff} and \ref{tab:nubareff} for
\nuA and \nubarA interactions, respectively.
The samples are  normalized in such a way that the efficiency at the first bin
is taken as  1.00.
Given the fact that Bjorken $\mathrm{y}$ distributions in neutrino and anti-neutrino interactions are different,  
the momentum distribution of the positive muon  from \numubar CC interactions is  harder than that of  the negative muon in \numu CC interactions.
Therefore, the reconstruction and location efficiencies of \numubar CC events are higher than that of \numu CC events.
The efficiency does not depend on the black and grey track multiplicity at the primary vertex.

In   \numu (\numubar) CC interactions, the full kinematics of the event  can be 
reconstructed from the measurement of the muon momentum $p_{\mu}$, the angle $\theta_{\mu}$ of the produced muon with 
respect to the beam 
axis, and $E_{had}$, the energy transfer to the hadronic system:

\begin{equation}
E_{\nu} = E_{\mu}+E_{had} \nonumber
\end{equation}
\begin{equation}
\Qsqnu = 2E_{\nu}(E_{\mu} - p_{\mu}~cos\theta_{\mu}) - m_{\mu}^{2}\nonumber
\end{equation}
\begin{equation}
\wsqnu = 2m_{n}(E_{\nu}-E_{\mu})+m_{n}^{2} -\Qsqnu \nonumber
\end{equation} 
where $E_{\nu}$ and  $E_{\mu}$ are the energy of the incoming (anti-)neutrino and  (anti-)muon energy respectively, 
-\Qsqnu
the squared four-momentum transfer, $m_{n}$ and $m_{\mu}$ are the mass of the nucleon and muon respectively, 
 \wsqnu is  
the square of the invariant mass of the hadronic system.
 The average neutrino energy of the neutrino sample
is  \meanEnu = (38.0 $\pm$ 1.4) \GeV,
the mean-square momentum transferred
to the hadronic system is \meanQsqnu = (8.6 $\pm$0.5) \GeVcsq, and
the mean-square of the invariant mass of the hadronic system
is \wsqnu = (26.2 $\pm$ 1.3) \GeVccsq.
For the antineutrino sample these values are
\meanEnubar = (42.5 $\pm$ 1.6) \GeV,
\meanQsqnubar = (4.6 $\pm$ 0.3) \GeVcsq, and
\wsqnubar = (17.7 $\pm$ 0.8) \GeVccsq, respectively.
All distributions presented hereafter include the  efficiency correction.

\section{Multiplicity Distributions}

\begin{figure}[tb]
\begin{center} \resizebox{0.75\textwidth}{!}{
\includegraphics{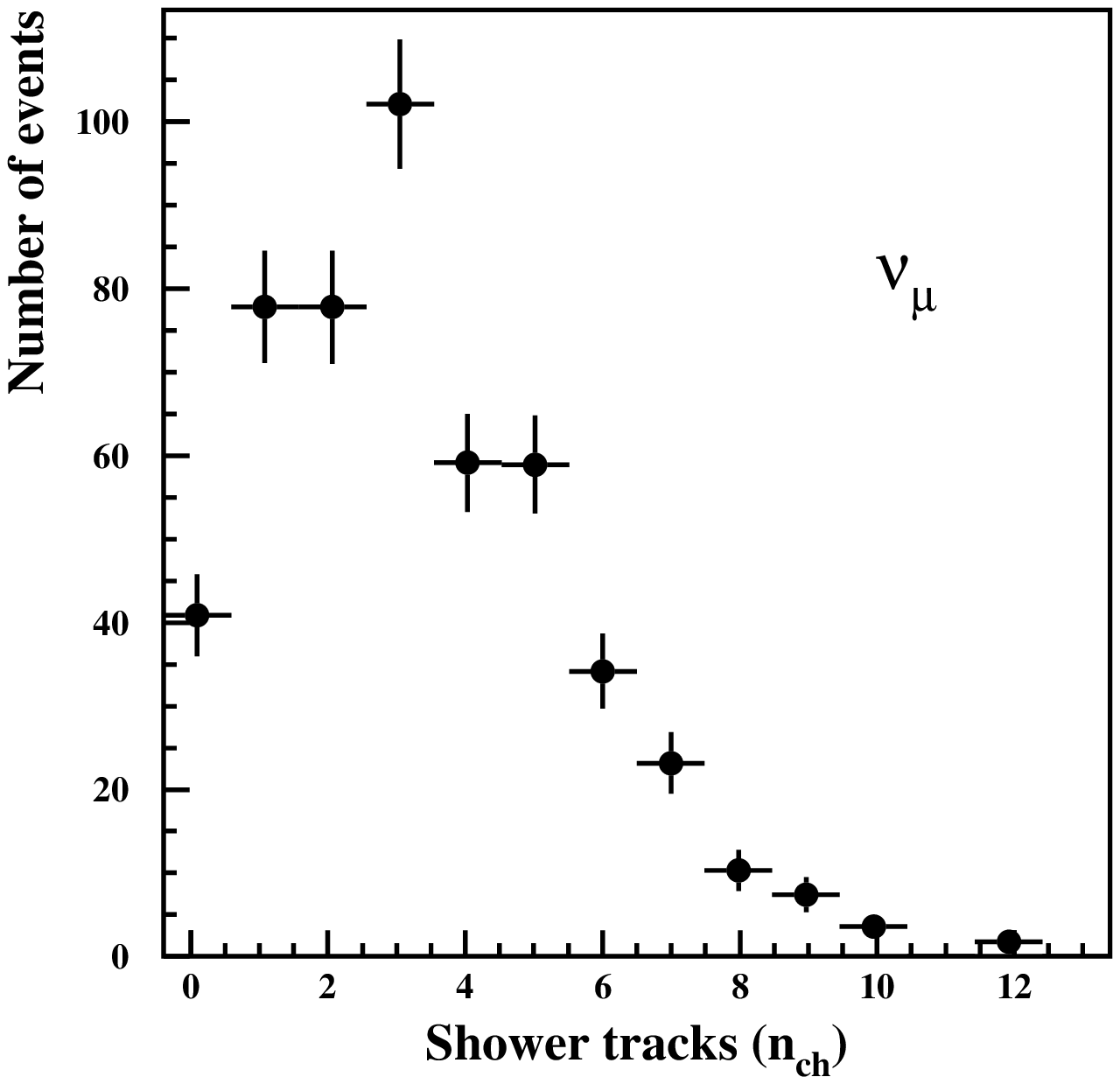}
\hspace{1cm} 
\includegraphics{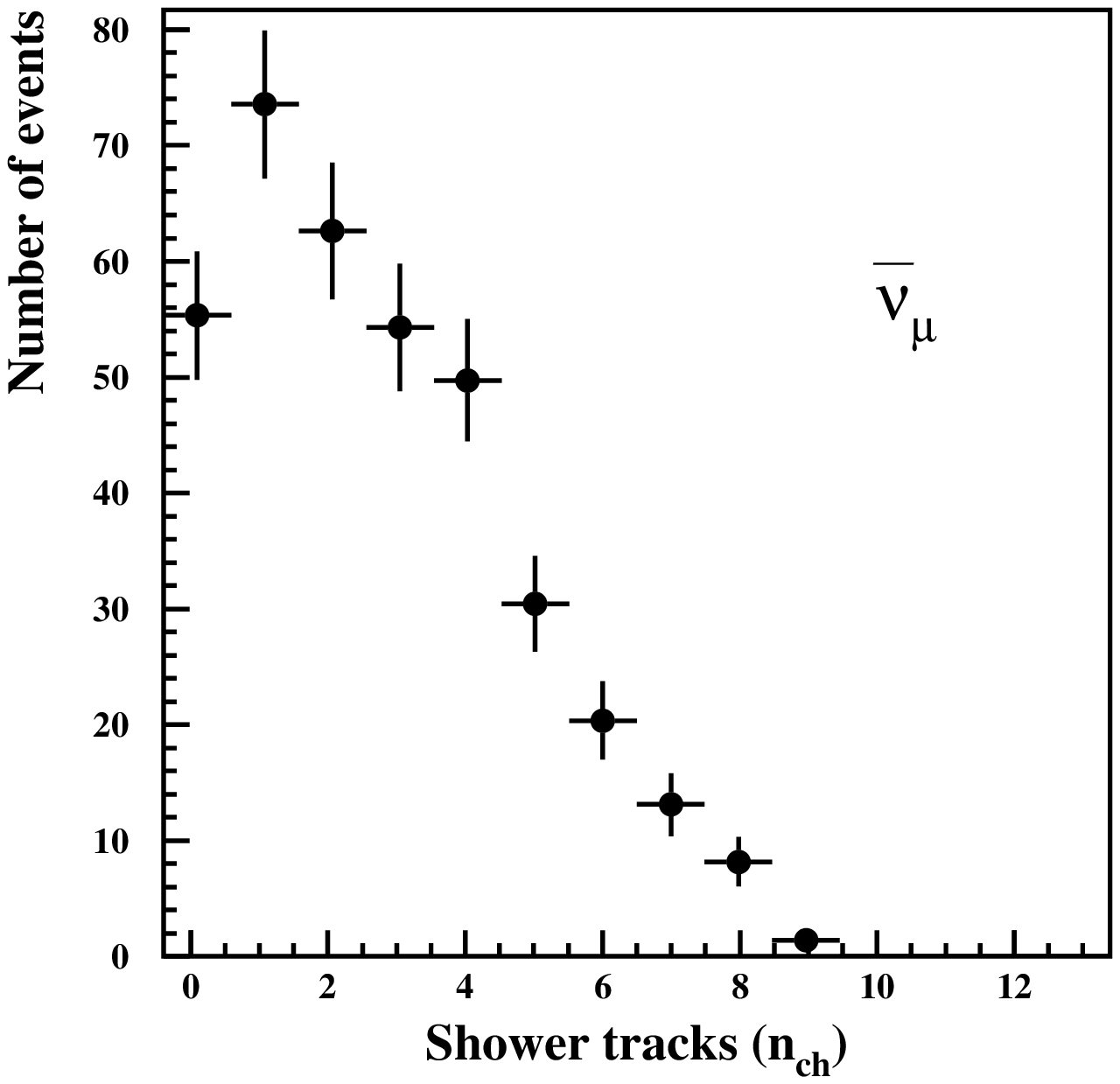}}
\end{center}
\begin{center} \resizebox{0.75\textwidth}{!}{
\includegraphics{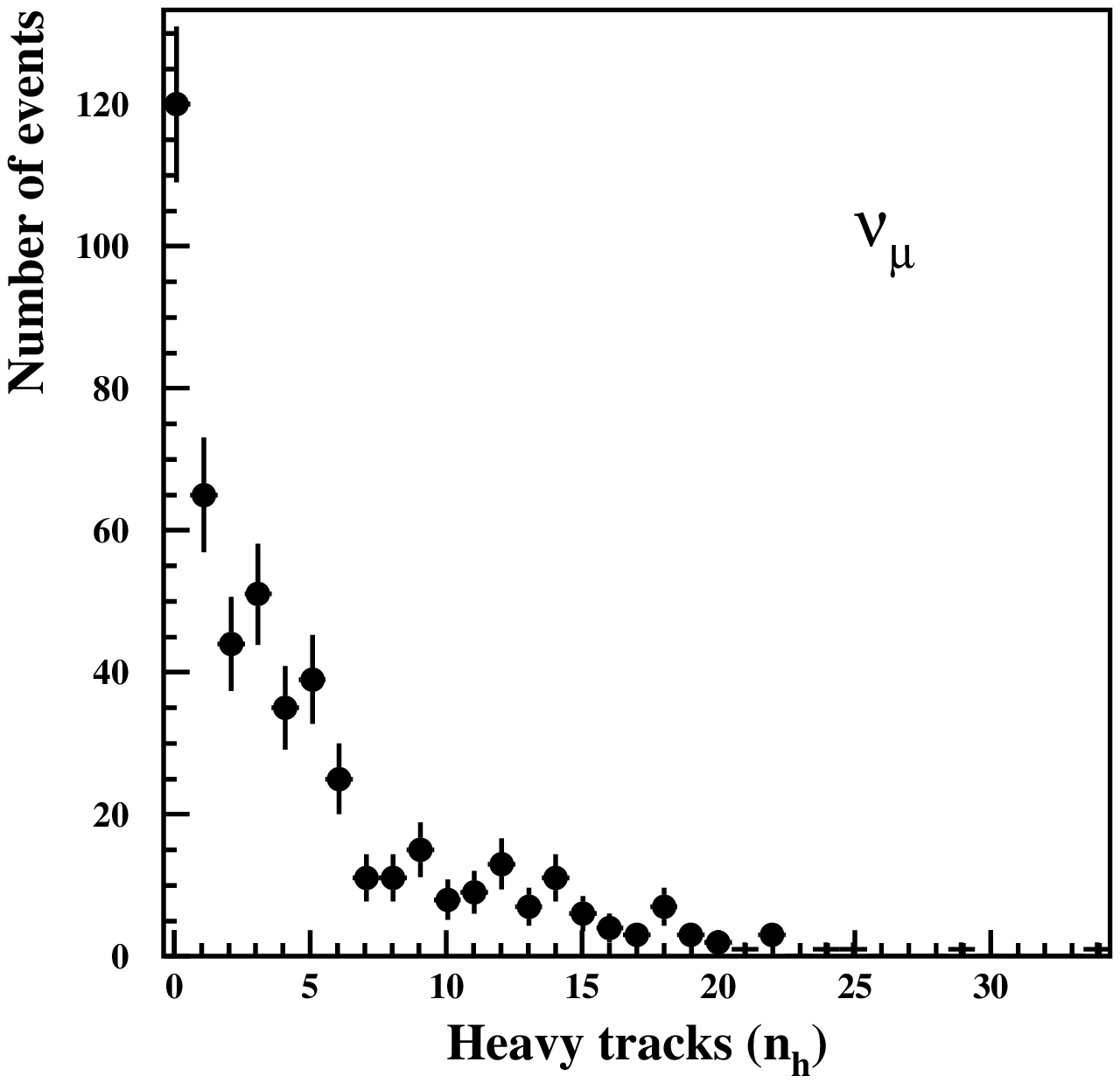}
\hspace{1cm} 
\includegraphics{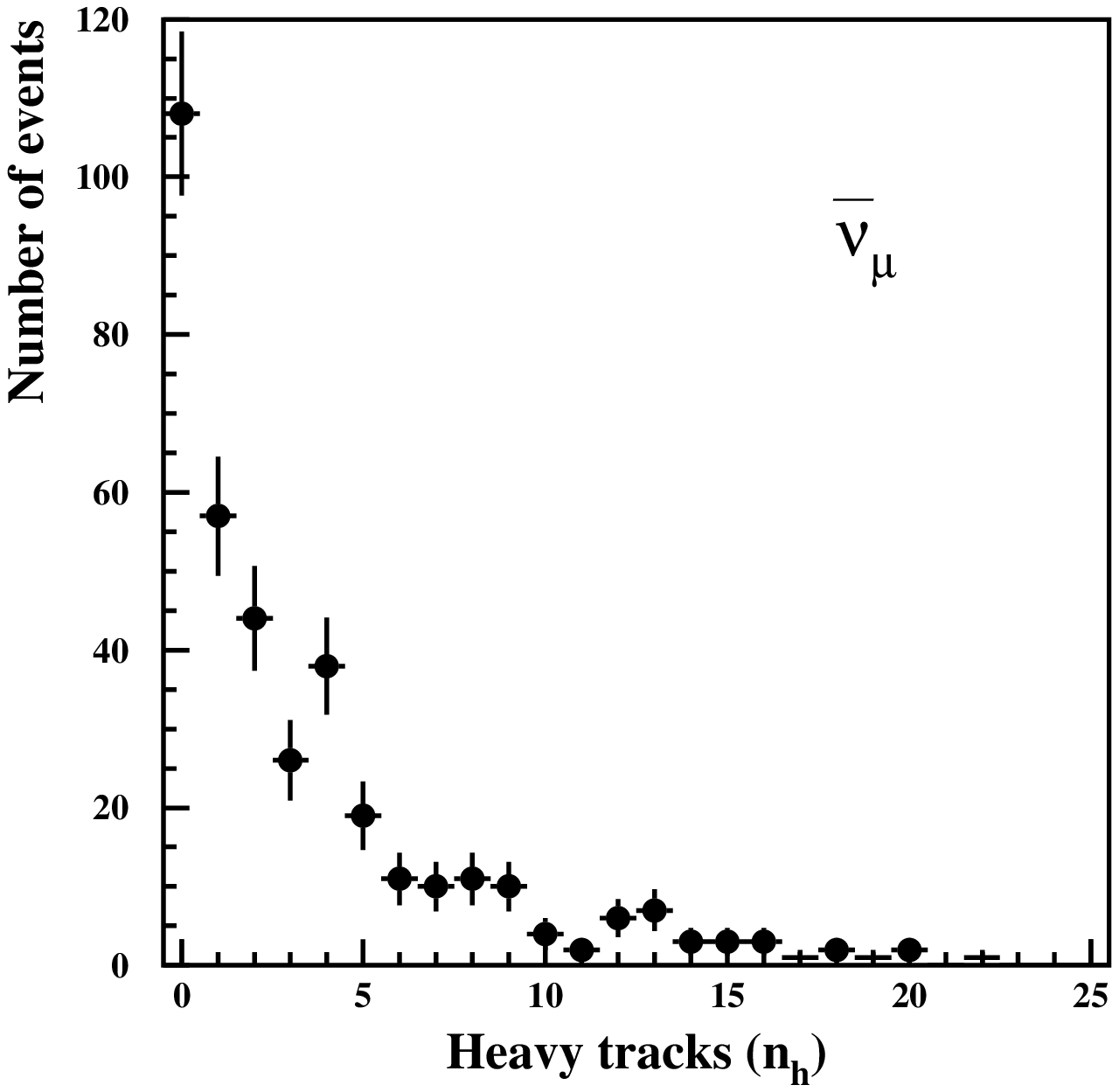}}
\caption{Multiplicity distributions for hadronic shower (top) and black plus grey (bottom) tracks in
{\numu}--nucleus and 
{\numubar}--nucleus interactions
\label{fig:mult}}
\end{center}
\end{figure}

The hadronic shower and heavy prong multiplicity
distributions based on the pseudo-rapidity selection are shown in Fig.~\ref{fig:mult}  and the
distribution as a function of emission angle of shower tracks is given in Table~\ref{tab:emiss}\footnote{The
corrected number of events is 1231(844) for \numu--A (\numubar--A) interactions}.
The average number of shower prongs in neutrino-nucleus interactions is
$\langle \nch(\nuA) \rangle = 3.4\pm{0.1}$.
The average heavy prong multiplicity is measured to be $\langle\nh(\nuA) \rangle =4.7\pm{0.2}$.
The average number of shower and heavy prongs in anti-neutrino induced
events are  measured to be
$\langle \nch(\nubarA) \rangle =2.8\pm{0.1} $ and
$\langle \nh(\nubarA) \rangle =3.5\pm{0.2}$, respectively.  These two quantities were never measured yet
in nuclear emulsion.
 
\begin{figure}[t]
\begin{center}
\vspace{1.5\baselineskip}
\resizebox{0.80\textwidth}{!}{
\includegraphics{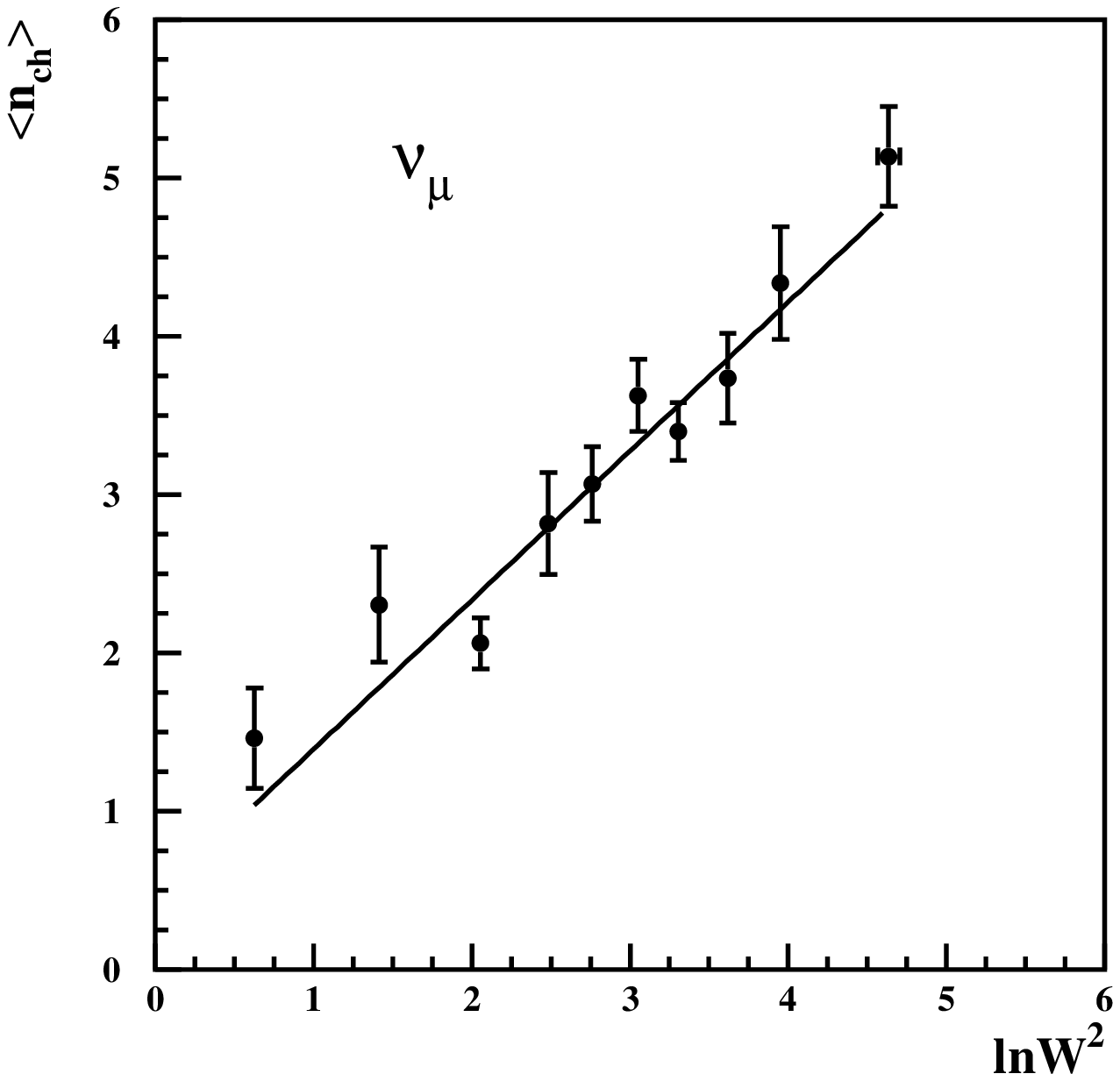}
\includegraphics{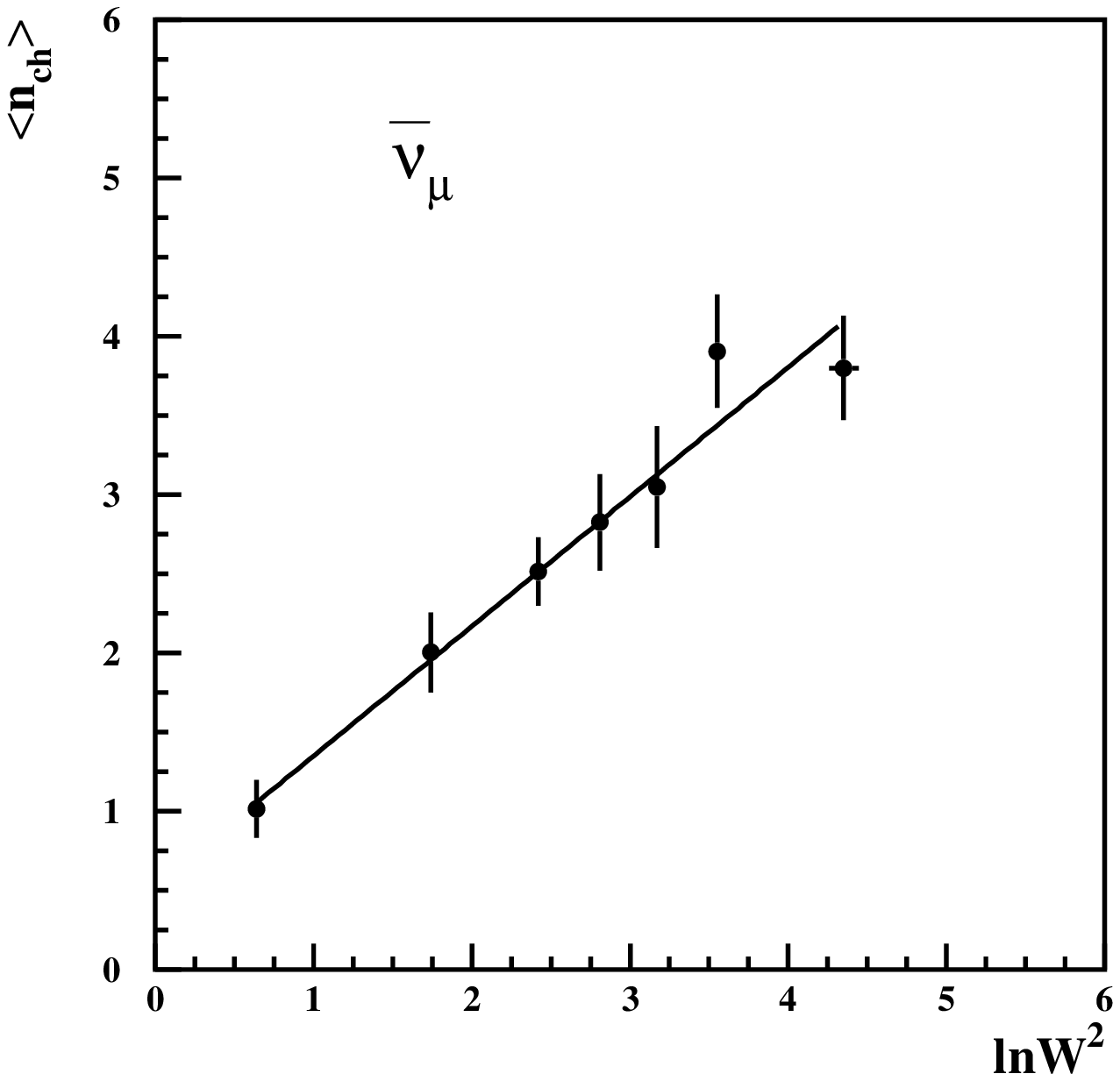}
}
\end{center}
\begin{center}
\vspace{1.0\baselineskip}
\caption{
The hadronic shower prong multiplicity distributions as a function of $\ln \wsq$ for 
\nuA, and  \nubarA interactions. 
\label{fig:lnw2}}
\end{center}
\end{figure}

The average shower prong multiplicities \nch for \nuA and
\nubarA interactions are plotted as a functions of
\wsq in Fig.~\ref{fig:lnw2}.
The numerical values  are given in Tables~\ref{tab:mult} and \ref{tab:mult2}.
The mean multiplicities are in  good agreement with a linear
dependence on
$\ln \wsq$,
\begin{equation}
\meannch = a + b \ln \wsq  \ .
\label{showerlinea}
\end{equation}
This expression was fitted to the average multiplicities for \wsq between
$1 \ \GeVccsq$ and $200 \ \GeVccsq$.
The values for the fit parameters $a$ and $b$ were found to be
\begin{equation}
\label{eq:nuab}
 \nch(\nuA) = (0.45\pm{0.24})+(0.94\pm{0.08}) \ln \wsq \nonumber
\end{equation}
\begin{equation}
\label{eq:nubarab}
\nch(\nubarA) = (0.53\pm{0.20})+(0.82\pm{0.08}) \ln \wsq . \nonumber
\end{equation}
The fitted values of $a$ and $b$  obtained in other experiments are shown in
Table~\ref{tab:comparison}. Our result on $a$ disagrees with the unpublished result of ~\cite{Voyvodik}
at  two sigma level. However we note that the value of $a$ found in
Ref~\cite{Voyvodik} is much larger than the corresponding value found in all other experiments.

\begin{sidewaystable}
\begin{center}
\caption{The  reconstruction and location  efficiency   in
  \nuA as a function of  the invariant mass of
the hadronic system  \wsq and the number of shower tracks.}
\vspace{0.5\baselineskip}
\label{tab:nueff}
\begin{tabular}{|c|c|c|c|c|c|c|c|c|c|c|c|c|c||}
\hline
 \wsq &
\multicolumn{7}{|c|}{$\nch = \nsh - 1$}\\ \cline{2-8}
 \GeVccsq  & 0 & 1 & 2 & 3 & 4 & $\ge$ 5 & All  \\ \hline

$1\div3$ &1.00$\pm$ 0.03 &0.95$\pm$ 0.01&0.98$\pm$ 0.01&0.88$\pm$ 0.04
&0.79$\pm$ 0.08&0.71$\pm$ 0.43 & 0.96$\pm$0.01\\

$3\div6$ &0.89$\pm$ 0.04 &0.92$\pm$ 0.01& 0.93$\pm$ 0.01&0.88$\pm$ 0.01
&0.92$\pm$ 0.04&0.88$\pm$ 0.09& 0.91$\pm$0.01\\

$6\div10$ &0.85$\pm$ 0.04&0.85$\pm$ 0.01&0.88$\pm$ 0.01 &0.82$\pm$ 0.01 &0.80$\pm$ 0.02
&0.73$\pm$ 0.04& 0.84$\pm$0.01\\

$10\div14$ & 0.76$\pm$ 0.07 &0.77$\pm$ 0.02 & 0.81$\pm$ 0.02 &0.78$\pm$ 0.01 &0.81$\pm$ 0.02
&0.73$\pm$ 0.02& 0.78$\pm$0.01\\

$14\div18$ &0.40$\pm$ 0.12 &0.75$\pm$ 0.02 & 0.76$\pm$ 0.02 &0.72$\pm$ 0.01 &0.78$\pm$ 0.02
&0.71$\pm$ 0.02& 0.74$\pm$0.01\\

$18\div24$ &0.47$\pm$ 0.12 &0.72$\pm$ 0.02 & 0.74$\pm$ 0.02 &0.71$\pm$ 0.01 &0.75$\pm$ 0.02
&0.72$\pm$ 0.02& 0.72$\pm$0.01\\

$24\div32$ & 0.31$\pm$ 0.15 &0.71$\pm$ 0.03 & 0.66$\pm$ 0.03 &0.67$\pm$ 0.01 &0.71$\pm$ 0.02
&0.69$\pm$ 0.01& 0.69$\pm$0.01\\

$32\div45$ & 0.55$\pm$ 0.24 &0.67$\pm$ 0.03 & 0.69$\pm$ 0.03 &0.68$\pm$ 0.02 &0.64$\pm$ 0.02
&0.66$\pm$ 0.01& 0.67$\pm$0.01\\

$45\div63$ & 0.43$\pm$ 0.35 &0.61$\pm$ 0.05 & 0.58$\pm$ 0.05 &0.66$\pm$ 0.02 &0.66$\pm$ 0.02
&0.60$\pm$ 0.01& 0.62$\pm$0.01\\

$>63$ & 0.77$\pm$ 0.28 &0.56$\pm$ 0.05 & 0.66$\pm$ 0.05 &0.59$\pm$ 0.02 &0.62$\pm$ 0.02
&0.53$\pm$ 0.01& 0.55$\pm$0.01\\
\hline
All  & 0.91$\pm$0.01 & 0.86$\pm$0.01 & 0.86$\pm$0.01 & 0.74$\pm$0.01 &0.73$\pm$0.01&0.63$\pm$0.01 & 0.75$\pm$0.01 \\
\hline
\end{tabular}
\end{center}
\begin{center}
\caption{The reconstruction and location  efficiency  in \nubarA as a function of the invariant mass of 
the hadronic system \wsq and the number of shower tracks.}
\vspace{0.5\baselineskip}
\label{tab:nubareff}
\begin{tabular}{|c|c|c|c|c|c|c|c|c|c|c|c||}
\hline
 \wsq &
\multicolumn{6}{|c|}{$\nch = \nsh - 1$}\\ \cline{2-7}
 \GeVccsq  & 0 & 1 & 2& 3 & $\ge$4 & All    \\ \hline

$1\div4$ & 1.00$\pm$ 0.02& 0.96$\pm$ 0.02& 0.98$\pm$ 0.02&0.99$\pm$ 0.05&0.92$\pm$ 0.19& 0.98$\pm$0.01\\

$4\div8$ & 0.95$\pm$ 0.03& 0.95$\pm$ 0.02& 0.95$\pm$ 0.02&0.96$\pm$ 0.04&0.88$\pm$ 0.06 & 0.94$\pm$0.01\\

$8\div14$ & 0.96$\pm$ 0.04& 0.90$\pm$ 0.03& 0.90$\pm$ 0.02&0.99$\pm$ 0.03&0.91$\pm$ 0.03& 0.92$\pm$0.01\\

$14\div20$ & 0.82$\pm$ 0.07& 0.67$\pm$ 0.05& 0.92$\pm$ 0.03&0.99$\pm$ 0.04&0.90$\pm$ 0.03& 0.92$\pm$0.01\\

$20\div28$ & 0.97$\pm$ 0.08& 0.73$\pm$ 0.06& 0.83$\pm$ 0.04&0.88$\pm$ 0.04&0.90$\pm$ 0.02& 0.86$\pm$0.01\\

$28\div45$ & 0.85$\pm$ 0.09& 0.79$\pm$ 0.05& 0.86$\pm$ 0.04&0.88$\pm$ 0.04&0.90$\pm$ 0.02& 0.87$\pm$0.01\\
$>45$ & 0.74$\pm$ 0.09& 0.65$\pm$ 0.05& 0.73$\pm$ 0.04&0.73$\pm$ 0.03&0.67$\pm$ 0.01& 0.68$\pm$0.01\\
\hline
All  & 0.97$\pm$ 0.01& 0.90$\pm$ 0.01& 0.92$\pm$ 0.01&0.91$\pm$ 0.01& 0.81 $\pm$0.01 & 0.90$\pm$0.01\\
\hline
\end{tabular}
\end{center}
\end{sidewaystable} 
\begin{table}[tb]
\caption{The number of shower hadrons  in \numu (\numubar) as a
function of emission angle $\theta$ where $\bf\mathrm{N_{tr}}$ is the number of hadron tracks 
${\bf\mathrm{\frac{N_{tr}}{N_{ev}}}}$ is the ratio of the number hadrons tracks to the total number of events.}
\label{tab:emiss}
\vspace{0.5\baselineskip}
\begin{center}
\begin{tabular}{cccc}
\hline
{\bf $\theta$ (Radian)} & {\bf $\meantheta$} &
{$\bf\mathrm{N_{tr}}$} & ${\bf\mathrm{\frac{N_{tr}}{N_{ev}}}}$\\
\hline
$ 0.000 \div 0.050 $ & 0.032$\pm$0.001 (0.031$\pm$0.001) & 346 (161) & 0.28(0.19)\\
$ 0.050 \div 0.100 $ & 0.075$\pm$0.001 (0.073$\pm$0.001) & 554 (282) & 0.45(0.34)\\
$ 0.100 \div 0.150 $ & 0.125$\pm$0.001 (0.123$\pm$0.001) & 510 (270) & 0.41(0.32)\\
$ 0.150 \div 0.200 $ & 0.174$\pm$0.001 (0.176$\pm$0.001) & 394 (280) & 0.32(0.33)\\
$ 0.200 \div 0.300 $ & 0.245$\pm$0.002 (0.246$\pm$0.002) & 774 (394) & 0.63(0.47)\\
$ 0.300 \div 0.400 $ & 0.350$\pm$0.002 (0.347$\pm$0.002) & 538 (330) & 0.44(0.39)\\
$ 0.400 \div 0.500 $ & 0.451$\pm$0.002 (0.449$\pm$0.003) & 380 (278) & 0.31(0.33)\\
$ 0.500 \div 0.600 $ & 0.544$\pm$0.003 (0.544$\pm$0.003) & 290 (171) & 0.24(0.20)\\
$       >    0.600 $ & 0.651$\pm$0.003 (0.655$\pm$0.004) & 210 (121) & 0.17(0.15)\\
\hline
\end{tabular}
\end{center}
\end{table}


\begin{sidewaystable}
\begin{center}
\caption{The shower multiplicity distributions in
  \nuA as a function of \wsq (errors shown are statistical only).}
\vspace{0.5\baselineskip}
\label{tab:mult}
\begin{tabular}{|c|c|c|c|c|c|c|c|c|c|c|c|c|c|c|c|c|c|}
\hline
\wsq & \meanwsq  & \meannch & \meannbl  &
\multicolumn{12}{|c|}{$\nch = \nsh - \mbox{muon}$} & {\bf Events} \\ \cline{5-17}
 \GeVccsq & \GeVccsq  & & & 0 & 1 & 2& 3 & 4 & 5 & 6 & 7 & 8 & 9 & 10 & $\ge$11 &  \\ \hline
$1\div3$ & 1.95$\pm$0.07 & 1.39$\pm$0.19 & 4.39$\pm$0.64 & 21 & 17 & 15 & 3 & 2 & 1 & 1 & 1 & 0 & 0 & 0 
& 0 & 61 \\
$3\div6$ & 4.18$\pm$0.11 & 2.21$\pm$0.27 & 5.80$\pm$0.93 & 10 & 12 & 12 & 10 & 5 & 1 & 2 & 0 & 0 & 2 & 
0 & 0 & 54\\
$6\div10$ & 7.89$\pm$0.14 & 2.08$\pm$0.19 & 3.76$\pm$0.59 & 9 & 17 & 10 & 18 & 4 & 1 & 2 & 0 & 0 & 0 & 
0 & 0 & 61\\
$10\div14$ & 12.09$\pm$0.18 & 2.86$\pm$0.29 & 3.28$\pm$0.61 & 3 & 13 & 12 & 7 & 6 & 4 & 2 & 2 & 0 & 0 & 
1 & 0 & 50\\
$14\div18$ & 15.88$\pm$0.18 & 3.22$\pm$0.26 & 4.42$\pm$0.67 & 3 & 10 & 7 & 13 & 6 & 6 & 5 & 3 & 0 & 0 & 
0 & 0 & 53\\
$18\div24$ & 21.22$\pm$0.23 & 3.63$\pm$0.26 & 4.18$\pm$0.71 & 1 & 5 & 6 & 19 & 5 & 6 & 4 & 5 & 1 & 0 & 
0 & 0 & 52\\
$24\div32$ & 27.37$\pm$0.32 & 3.37$\pm$0.21 & 3.38$\pm$0.57 & 0 & 6 & 11 & 12 & 10 & 8 & 5 & 1 & 0 & 0 
& 0 & 0 & 53\\
$32\div45$ & 37.36$\pm$0.59 & 3.67$\pm$0.34 & 3.10$\pm$0.69 & 1 & 4 & 6 & 6 & 6 & 6 & 2 & 2 & 2 & 0 & 0 
& 0 & 35 \\
$45\div63$ & 52.26$\pm$0.32 & 4.43$\pm$0.36 & 5.13$\pm$0.89 & 0 & 2 & 6 & 7 & 9 & 7 & 2 & 2 & 2 & 1 & 0 
& 1 & 39 \\
$>63$ & 112.50$\pm$7.78 & 4.95$\pm$0.36 & 3.43$\pm$0.53 & 0 & 2 & 3 & 5 & 4 & 12 & 4 & 2 & 3 & 2 
& 1 & 
0 & 38 \\
\hline
Total & \multicolumn{15}{c}{~} &     496 \\ \hline
\end{tabular}
\end{center}
\end{sidewaystable}
\begin{sidewaystable}
\begin{center}
\caption{The shower multiplicity distributions in
  \nubarA as a function of  \wsq (errors shown are statistical only).}
\vspace{0.5\baselineskip}
\label{tab:mult2}
\begin{tabular}{|c|c|c|c|c|c|c|c|c|c|c|c|c|c|c|c|c|c|c|c||}
\hline
 \wsq & \meanwsq  &\meannch  & \meannbl & 
\multicolumn{10}{|c|}{$\nch = \nsh - \mbox{muon}$} & {\bf Events} \\ \cline{5-15}
 \GeVccsq & \GeVccsq  & & & 0 & 1 & 2 & 3 & 4 & 5 & 6 & 7 & 8 &$\ge$ 9 & \\ \hline
$1\div4$  & 2.06$\pm$0.10 & 1.02$\pm$0.17 & 3.53${\pm}$0.51 & 31 & 21 & 5  & 3  & 3  & 1 & 1 & 0 & 0 & 0 &  65 \\

$4\div8$   & 5.84$\pm$0.14 & 1.94$\pm$0.20 & 3.21$\pm$0.48 & 15 & 13 & 15 & 9  & 10 & 2 & 0 & 0 & 1 & 0 &  65 \\

$8\div14$  & 11.38$\pm$0.24 & 2.58$\pm$0.22 & 4.25$\pm$0.65 & 5 & 12  & 15 & 11 & 9  & 3 & 0 & 3 & 0 & 0 &  58 \\

$14\div20$ & 16.65$\pm$0.22 & 2.96$\pm$0.26 & 2.99$\pm$0.54 & 3 & 11  & 13 & 10 & 3  & 4 & 8 & 1 & 0 & 0 &  53 \\

$20\div28$ & 23.97$\pm$0.30 & 3.04$\pm$0.31 & 3.24$\pm$0.55 & 4 & 9   & 12 &  9 & 4  & 5 & 3 & 3 & 1 & 1 &  51 \\

$28\div45$ & 35.06$\pm$0.72 & 3.89$\pm$0.29 & 2.42$\pm$0.49 & 2 & 4   & 4  &  8 & 10 & 8 & 4 & 3 & 2 & 0 &  45 \\

$>45$      & 87.73$\pm$8.34 & 3.73$\pm$0.35 & 2.87$\pm$0.71 & 1 & 5   & 1  &  6 & 9  & 5 & 2 & 1 & 2 & 0 &  32 \\
\hline
Total & \multicolumn{13}{c}{~} & 369 \\ \hline
\end{tabular}
\end{center}
\end{sidewaystable}

\begin{table}[tb]
\caption{The values of the parameters $a$ and $b$ obtained by fitting  the relation $\nch = a + b \ln \wsq$ 
to the
average charged hadron multiplicity $\nch$. The results of other neutrino experiments are
shown for comparison.
`Em' stands for nuclear emulsion; \meannpls and \meannneg
are the mean multiplicities of positively and negative charged particles.}
\label{tab:comparison}
\vspace{0.5\baselineskip}
\begin{center}
\begin{tabular}{ccccc}
\hline
{\bf Reaction} & {\bf \Enu (\GeV)} & {\bf a} &
{\bf b} & {\bf Ref.}\\
\hline
\numu -- Em & 40 & 0.45 $\pm$ 0.24 & 0.94 $\pm$ 0.08& This paper \\
\numu -- Em & 50 & 1.92$\pm$ 0.68 & 1.19$\pm$ 0.23 & \cite{Voyvodik} \\
\numu -- Em & 8.7& 1.07 $\pm$ 0.05 & 1.32$\pm$0.11 & \cite{Ammosov}\\
\numu -- p  & -& 0.37 $\pm$ 0.02 & 1.33 $\pm$ 0.02 & \cite{Allen}\\
$\nu$ -- Freon(CF$_{3}$Br) & & & \\
\meannpls       &  6 &0.15 $\pm$ 0.09 & 0.84$\pm$ 0.05 & \cite{Baranov}\\
\meannneg       &  6 &-0.49 $\pm$ 0.06 & 0.63 $\pm$ 0.04 &\cite{Baranov}\\
\numubar -- Em  & 40 & 0.53 $\pm$ 0.20 & 0.82 $\pm$ 0.08& This paper\\
\numubar -- p   & - &0.06 $\pm$ 0.06 & 1.22 $\pm$ 0.03 & \cite{Derrick}\\
\numubar -- Freon &&& \\
\meannpls      &  6 & -0.26 $\pm$ 0.20 & 0.56$\pm$ 0.12 &\cite{Baranov}\\
\meannneg      &  6 & 0.16 $\pm$ 0.20 & 0.52 $\pm$ 0.11 & \cite{Baranov}\\
\numubar -- n  & - & 0.80 $\pm$ 0.09 & 0.95$\pm$ 0.04 &\cite{Barlag}\\
\numubar -- p  &  - & 0.02 $\pm$ 0.20 & 1.28 $\pm$ 0.08 & \cite{Barlag}\\
\hline
\end{tabular}
\end{center}
\end{table}
\begin{table}[h]
\caption{(QE+RES)-like events in \nuA and \nubarA interactions.}
\label{tab:quasi}
\vspace{0.5\baselineskip}
\begin{center}   
{\small
\begin{tabular}{ccccc}
\hline
         & $\ngr$       & 0    & 1  & {\bf Total} \\ \hline
\numu    & \nch = 0     & 39   & 39 & 78  \\ 
\numu    & \nch = 1     & 56   & 23 & 79  \\
\numubar & \nch = 0     & 132  & 25 & 157  \\
\numubar & \nch = 1     & 44   & ~  &  44    \\ 
\end{tabular}}
\end{center}
\end{table} 

For the first time a sample of (anti-)neutrino events measured in emulsions is large enough to study 
(QE+RES)-like topologies. In order to have a minimum bias sample of
(QE+RES)-like events, the
$\wsq\ge 1$ \GeVccsq cut was not applied to  the located events. Hence,  the starting sample of
\numu--A (\numubar--A) interactions becomes 627 (529).
An event is defined
as being (QE+RES)-like,   if the number of shower prongs is zero or one and the number grey prongs zero or one 
for for \nuA interactions regardless of the
 number of black tracks. In order to obtain (QE+RES)-like enriched sample in \nubarA  interactions, the sum of shower prongs and grey prongs 
is required to be one or zero  regardless of the
 number of black tracks. The efficiency of this selection is estimated from the MC simulation to be 72\% (81\%) for \numu--A 
(\numubar--A) 
interactions. In order to suppress background from DIS interactions we further require $\wsq< 10$ \GeVccsq for both samples.
After this selection the number of (QE+RES)-like events is reported in~Table~\ref{tab:quasi}.
The average values of kinematical parameters, \meanEnu, \meanQsqnu, \meanwsq,
corresponding to these events are given in~Table~\ref{tab:quasikine}\footnote{Due to detector resolutions, there are 62(52)
events with \wsq$<$0  in neutrino(anti-neutrino) sample.  For 
these events, \wsq  is taken to be $m_{n}^2$. }. The 
number of background events that mimic (QE+RES)-like topology 
is obtained from the MC simulation to be 83.4 and 73.5 events  for  \numu--A and \numubar--A   interactions 
respectively. The ratio of  reconstruction and location efficiency of (QE+RES)-like  events to that of all CC events is found to be 1.22 (1.13) for  
\numu--A (\numubar--A) interactions.

After applying the efficiency  and background corrections,  the fraction of 
(QE+RES)-like events is  found to be
(13.4$\pm$1.0$\pm$2.0)$\%$ for \numu  and  (26.3$\pm$1.4$\pm$3.9)$\%$ for \numubar
interactions, respectively. These values are in reasonable agreement with the ones 
used in the Monte Carlo simulation~(Section 5). We have accounted for a systematic uncertainty of 15\% coming 
from the efficiency and background estimation by Monte Carlo modeling.

\begin{table}[h]
\caption{Kinematic variables for (QE+RES)-like events for \numu and \numubar.}
\label{tab:quasikine}
\vspace{0.5\baselineskip}
\begin{center}
{\small
\begin{tabular}{cccc}
\hline
             & \meanEnu (\GeV) & \meanwsq (\GeVccsq) & \meanQsq (\GeVcsq) \\ \hline
  \numu   \nch = 0, \ngr = 0  & 21.7 $\pm$ 2.2 & 1.9 $\pm$ 0.3   & 0.8 $\pm$ 0.1 \\
  \numu   \nch = 0, \ngr = 1  & 21.2 $\pm$ 1.9 & 2.6 $\pm$ 0.4   & 1.4 $\pm$ 0.2 \\
  \numu   \nch = 1, \ngr = 0  & 21.7 $\pm$ 2.0 & 2.9 $\pm$ 0.4   & 1.5 $\pm$ 0.2 \\ 
  \numu   \nch = 1, \ngr = 1  & 23.2 $\pm$ 2.7 & 3.1 $\pm$ 0.6   & 2.4 $\pm$ 0.3 \\ 
  \numubar \nch = 0, \ngr = 0 & 21.9 $\pm$ 2.2 & 1.9 $\pm$ 0.3   & 0.8 $\pm$ 0.1 \\
  \numubar \nch = 0, \ngr = 1 & 21.8 $\pm$ 2.0 & 2.4 $\pm$ 0.4   & 1.4 $\pm$ 0.2 \\
  \numubar \nch = 1, \ngr = 0 & 21.1 $\pm$ 2.0 & 3.8 $\pm$ 0.4   & 1.6 $\pm$ 0.2 \\

\end{tabular}}
\end{center}
\end{table}

The sub-sample of (QE+RES)-like \numu (\numubar) events with neither black nor grey prongs is important
for the understanding of nuclear mechanisms involving  hadrons in the nucleus.
The fraction of this type of topology is measured as 
(1.2$\pm$ 0.4$\pm$0.2)$\%$ for \numu and
(9.5$\pm$1.0$\pm$1.4)$\%$ for \numubar interactions, respectively. For  the neutrino  events the proton in the final 
state is absorbed by the nucleus without any visible activity. On the other hand, this fraction is larger for
anti-neutrinos since in the final state there is a neutron.

\section{Dispersion and Koba--Nielsen--Olesen distributions}

We have also investigated the dependence of the dispersion, \dch, of the shower hadron multiplicity distribution, 
 on the average multiplicity. The dispersion \dch is defined as
\begin{equation}
\dch \stackrel{\rm{def}}{=}\sqrt{\langle \nch^{2}\rangle-\langle
\nch \rangle^{2}} \ . 
\label{dispersions}
\end{equation}
For independent particle production, the multiplicity follows a Poisson
distribution. 
Therefore, one gets
\begin{equation}
\label{disp}
\dch = \sqrt{\meannch}.
\end{equation}
For charged particle production in hadronic interactions, however,
an empirical relation~\cite{Wroblewski} was found 
\begin{equation}
\dch = A + B \meannch \ .
\label{disper}
\end{equation}

\begin{figure}[tb]
\begin{center} \resizebox{0.80\textwidth}{!}{
\includegraphics{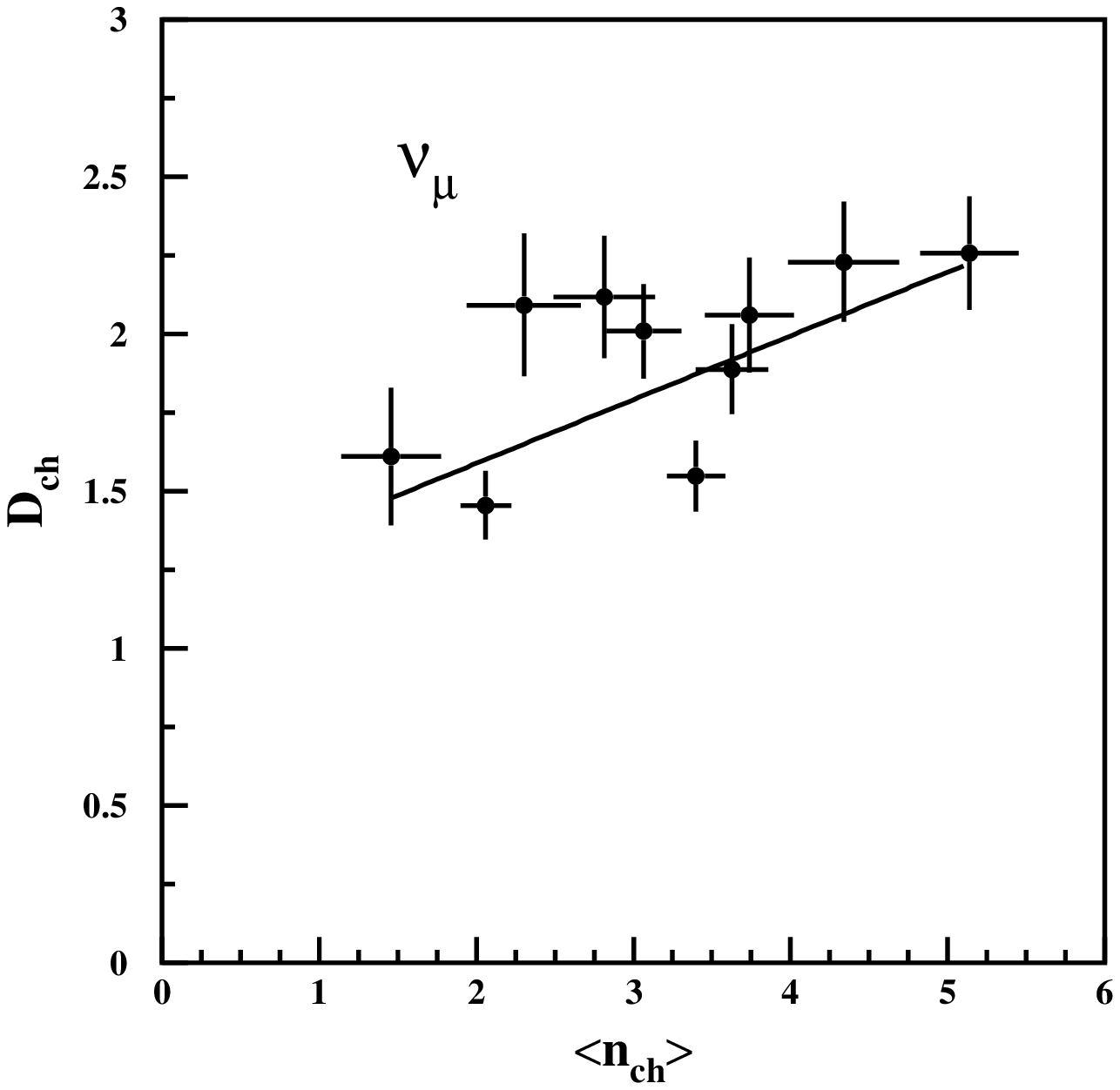}
\includegraphics{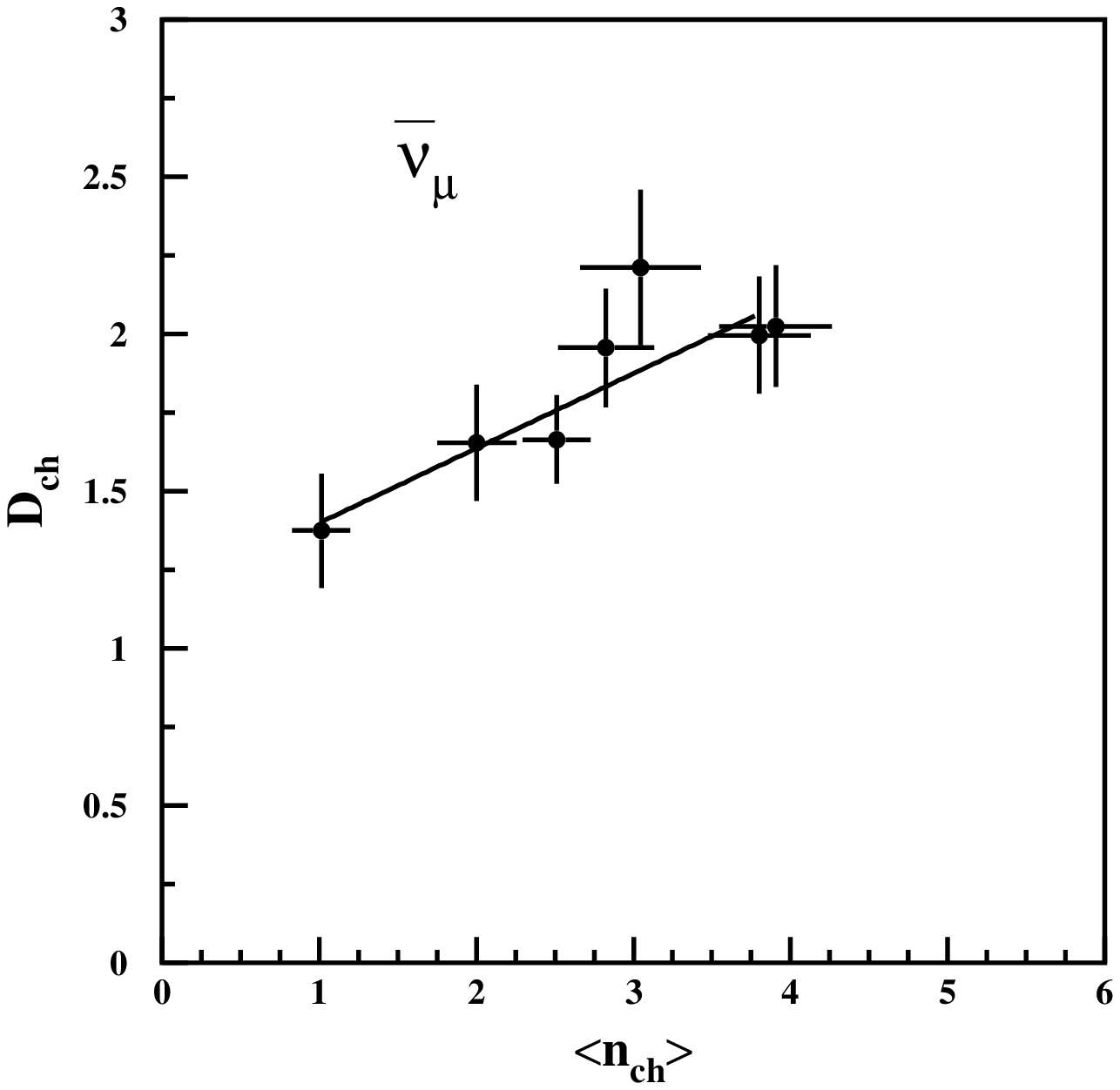}
}
\caption{
  The hadronic shower prong multiplicity dispersion as a function
  of  \meannch  for \nuA and  \nubarA interactions.
}
\label{fig:dispe}
\end{center}
\end{figure}
 
\begin{table}[tb]
\caption{Parameters of the linear fit to the dispersion
  distributions. The results of other neutrino experiments are also
  shown. 
  Note that $D_{+}$ and $D_{-}$ are the dispersion for positive and
  negative charged particles of reaction \nuA and \nubarA,
  respectively.} 
\label{tab:disptable}
\vspace{0.5\baselineskip}
\begin{center}

\begin{tabular}{ccccc}
\hline
{\bf Reaction} & {\bf A} & {\bf B} & {\bf Ref.}\\             
\hline 
\numu -- Em    & 1.18 $\pm$ 0.17 & 0.20 $\pm$0.05& This paper \\
\numu -- p     & 0.36 $\pm$ 0.03 & 0.36 $\pm$0.03 & \cite{Allen}\\
\numu -- Freon & & & \cite{Baranov}    \\
$D_{+}$        & 0.29 $\pm$ 0.12 & 0.36$\pm$0.06 &   \\
$D_{-}$        & 0.33 $\pm$ 0.03 & 0.50 $\pm$0.05 &  \\
\numubar -- Em     & 1.16 $\pm$ 0.21 & 0.24 $\pm$0.08& This paper\\
\numubar -- Freon  &&& \cite{Baranov} \\
$D_{+}$            &  0.30 $\pm$ 0.11 & 0.64$\pm$0.20 &  \\ 
$D_{-}$            &  0.35 $\pm$ 0.04 & 0.35 $\pm$0.06 &  \\
\numubar -- n      & -0.10 $\pm$ 0.04 & 0.42$\pm$0.02 & \cite{Barlag} \\ 
\numubar -- p      &  0.28 $\pm$ 0.06 & 0.36$\pm$0.04 & \cite{Barlag}\\
\hline 
\end{tabular} 
\end{center} 
\end{table} 

\begin{figure}[b]
\begin{center} \resizebox{0.75\textwidth}{!}{
\includegraphics{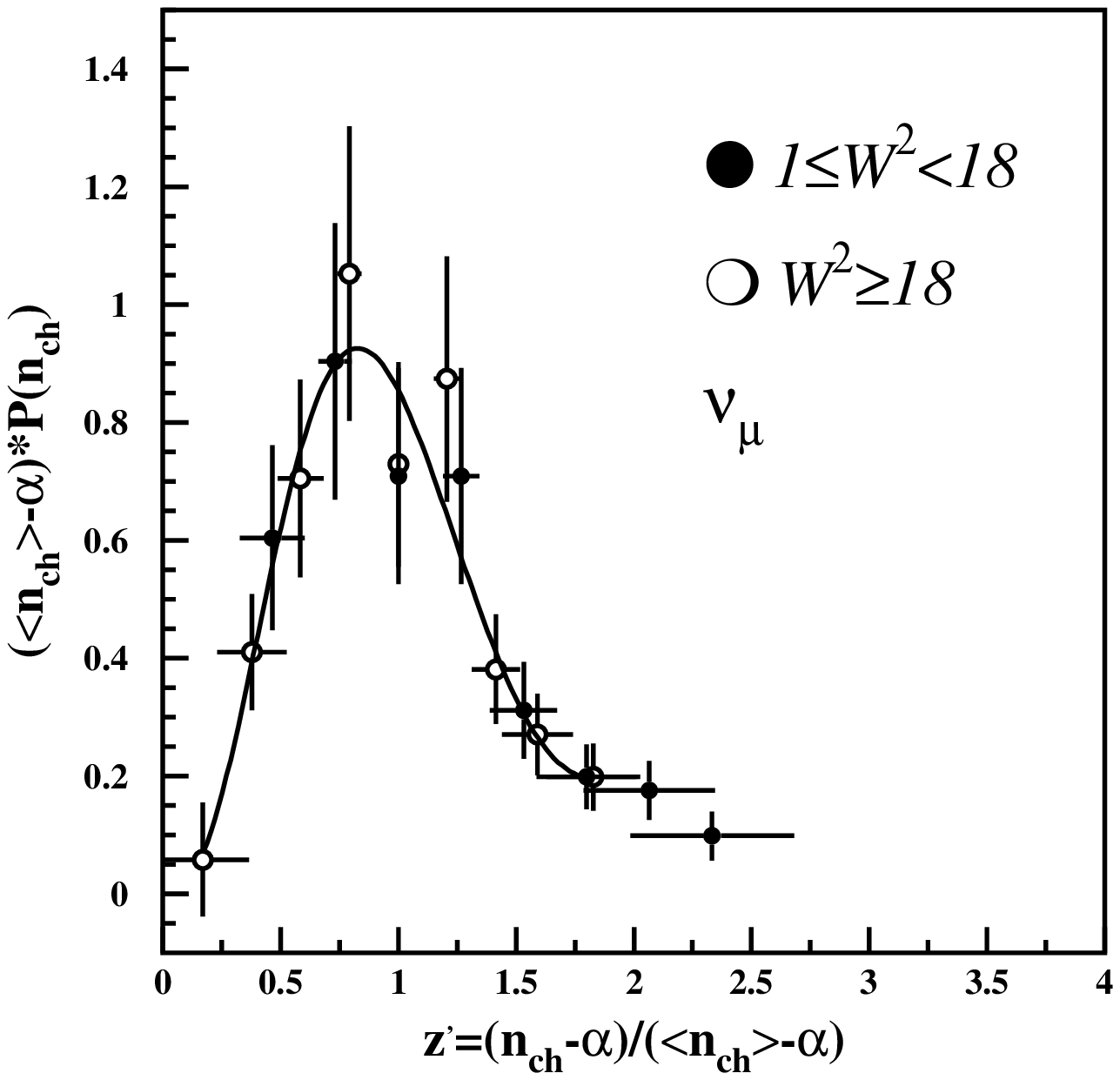}
\includegraphics{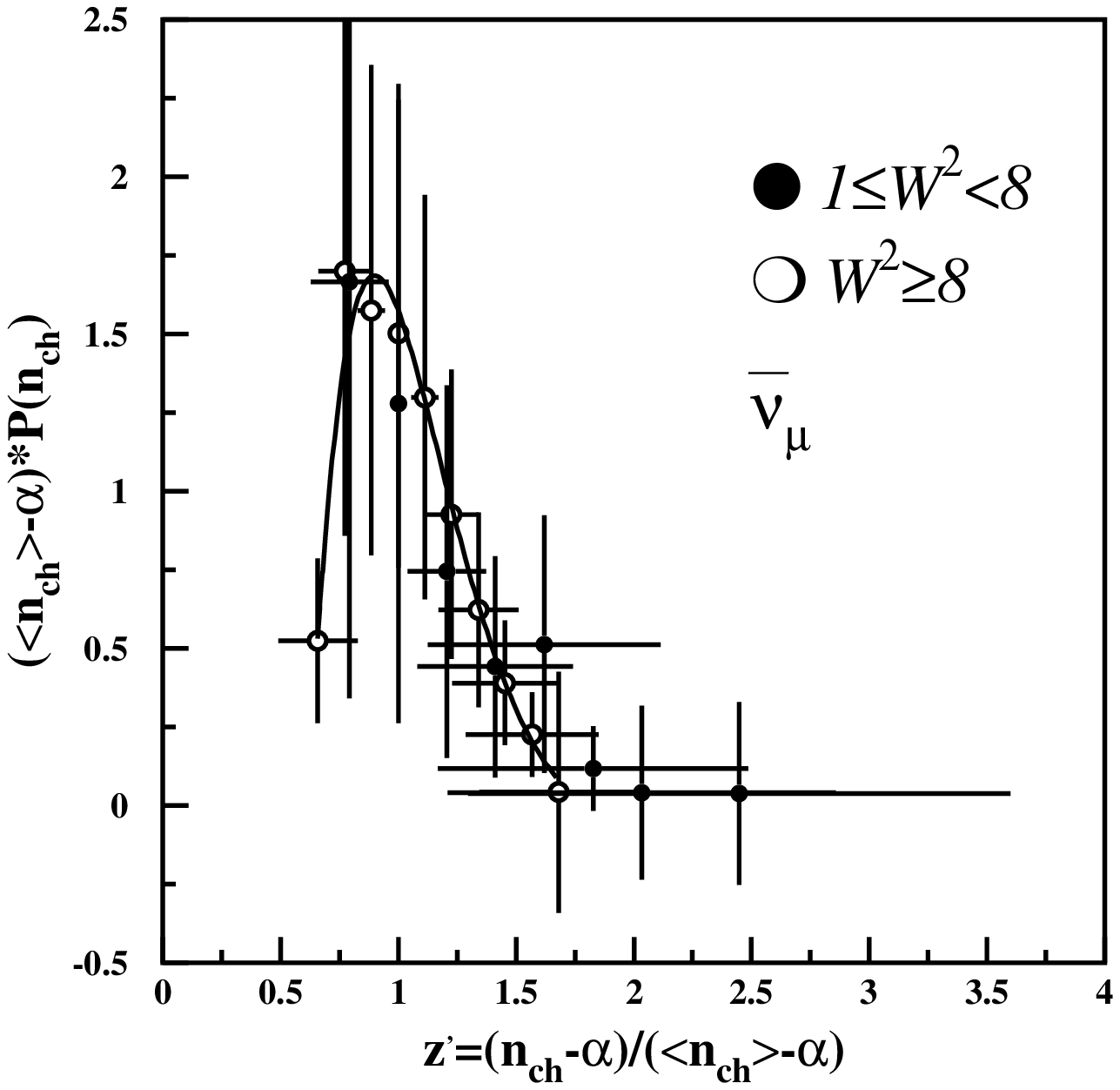}
}
\caption{The KNO scaling  distribution for shower prongs. The superimposed curve represents a fit to $\mathrm{p}{\mathrm{p}}$ 
data~\cite{slat}. The data approximately agree with KNO scaling, i.e., the data points at different \wsq intervals lie approximately 
on a single curve.}
\label{fig:kno}
\end{center}
\end{figure}

The linear dependence of the \dch on \meannch would imply that the shape of
the distribution of the normalized multiplicity
is independent of the hadronic effective mass $W$. 
Koba, Nielsen and
Olesen have shown that at asymptotically high energies, a
probability  $P(\nch)$ to produce an interaction of 
multiplicity \nch scaled by the average \meannch is independent of the energy.
\begin{equation}\label{knoeq}
  \meannch \cdot{P}(\nch) \ 
  \stackrel{E\rightarrow\infty}{\longrightarrow} \ 
  \Psi\left(z = \frac{\nch}{\meannch} \right) \ ,
\end{equation}
where $\Psi$ is the probability distribution  independent of energy.
This is called Koba-Nielsen-Olesen (KNO) scaling.
Exact KNO-scaling demands that, in the
linear dependence of the dispersion on the average multiplicity,
Eq.~\ref{disper}, the intercept parameter $A$ is exactly equal to
zero. For our event samples,  \dch is given as a function of \meannch in Fig.~\ref{fig:dispe} 
with a linear fit 
superimposed.
The values of the fit parameters are
\begin{equation}
 \dch(\nuA)  =  (1.18\pm{0.17})+(0.20\pm{0.05})\meannch \nonumber 
\end{equation}
\begin{equation}
 \dch(\nubarA)  =  (1.16\pm{0.21})+(0.24\pm{0.08})\meannch \nonumber
\end{equation}

The fitted values of $A$ and $B$   obtained  in this and  other
experiments are shown in Table~\ref{tab:disptable}. 
One can observe that our data, as well as freon data,
are not compatible with $A = 0$.  Indeed, $A = 0$ is expected only for a proton or neutron target. 
A non zero
value of $A$ is due to the heavy nuclear targets in nuclear emulsions.  
In such a  case one introduces  a new variable
$z'$ defined as~\cite{Baranov}
\begin{equation}\label{zeq}
  z' = \frac{\nch-\alpha}{\langle \nch-\alpha \rangle}.
\end{equation}

Here $\alpha = -\langle n_{0} \rangle$ and $\langle n_{0} \rangle$
is the extrapolated point where the fitted dispersion line crosses the
average multiplicity axis and found to be  -5.82  and -4.83 for
\nuA and \nubarA interactions, respectively.
The KNO-scaling law now reads
\begin{equation}\label{knocorr}
  K' = (\meannch - \alpha)P(\nch) = \Psi(z').
\end{equation}
Clearly, in the asymptotic limit it makes no difference whether one
uses $z$ or $z'$; in fact this expression describes an approach to restore KNO
scaling. 
Figure~\ref{fig:kno} shows the distributions obtained for $\Psi(z')$ as a
function of $z'$ for two intervals of \wsq, for  \nuA and \nubarA  samples; they are compatible with KNO scaling.
Similar results have been obtained in other
experiments~\cite{Hebert,Hebert2,Shivpuri}, however, with  
different values of $\alpha$. 

\section{Conclusion}

The multiplicity features  of (anti-)neutrino--nucleus interactions in emulsion
have been investigated. The results presented in this paper 
have been obtained with the main objective to aid in tuning 
(anti-)neutrino--nucleus interaction models of the Monte Carlo event
generators. 
To ease the use of these results, the numbers are presented in detail in
the form of tables.
The results can be summarized as follows.

For the first time with nuclear emulsion, the charged particle (shower, 
grey, black tracks)
production at the neutrino interaction vertex of (QE+RES)-like events has been studied. 
It has been found that  in  (1.2+-0.4+-0.2)
only the muon track is seen in the final state. These interactions are mainly 
from the reaction $\nu_\mu n\rightarrow \mu^{-} p$ and the proton is absorbed by the 
 nucleus without any visible recoil.
We also report the first study of anti-neutrino induced events
in nuclear emulsion for the production of charged particles.
The average number of shower and heavy prongs in anti-neutrino induced
events are  measured to be
$\langle \nch(\nubarA) \rangle =2.8\pm{0.1} $ and
$\langle \nh(\nubarA) \rangle =3.5\pm{0.2}$, respectively. 
The fraction of (QE+RES)-like events in anti-neutrino interactions  is  found to be
(26.3$\pm$1.4$\pm$3.9)$\%$.

The dependence of the average multiplicity \meannch   on $\ln{W}^{2}$ for \nuA and \nubarA  interactions is
compatible with being linear with similar slopes.
The dispersion \dch of the multiplicity distribution shows a linear
dependence on mean multiplicity \meannch.
The emulsion data are  consistent with the KNO scaling as a function of an appropriate 
multiplicity variable  
$z'$. 

\section*{Acknowledgements}
We gratefully acknowledge the help and support of the neutrino beam
staff and of the numerous technical collaborators who contributed to
the detector construction, operation, emulsion pouring, development,
and scanning.  The experiment has been made possible by grants from
the Institut Interuniversitaire des Sciences Nucl\'eair\-es and the
Interuniversitair Instituut voor Kernwetenschappen (Belgium), the
Israel Science Foundation (grant 328/94) and the Technion Vice
President Fund for the Promotion of Research (Israel), CERN (Geneva,
Switzerland), the German Bundesministerium f\"ur Bildung und Forschung
(Germany), the Institute of Theoretical and Experimental Physics
(Moscow, Russia), the Istituto Na\-zio\-na\-le di Fisica Nucleare
(Italy), the Promotion and Mutual Aid Corporation for Private Schools
of Japan and Japan Society for the Promotion of Science (Japan), the
Korea Research Foundation Grant (KRF-2003-005-C00014) (Republic of
Korea), the Foundation for Fundamental Research on Matter FOM and the
National Scientific Research Organization NWO (The Neth\-er\-lands),
and the Scientific and Technical Research Council of Turkey
(Turkey). We gratefully acknowledge their support.

\end{document}